\documentclass[letterpaper,preprint,aps]{revtex4-1}
\usepackage[latin9]{inputenc}
\usepackage{verbatim}
\usepackage{amsmath}
\usepackage{amssymb}
\usepackage{graphicx}

\makeatletter

\pdfpageheight\paperheight
\pdfpagewidth\paperwidth

\usepackage{graphicx}
\usepackage{tikz}
\usepackage{comment}

\DeclareMathOperator*{\var}{var}
\def\dd{{\rm d}}

\makeatother

\begin{document}
\title{Holographic Map for Cosmological Horizons}
\author{Chang Liu}
\email{chang\_liu3@brown.edu}

\author{David A. Lowe}
\email{lowe@brown.edu}

\affiliation{Department of Physics, Brown University, Providence, RI, 02912, USA}
\begin{abstract}
We propose a holographic map between Einstein gravity coupled to matter
in a de Sitter background and large $N$ quantum mechanics of a system
of spins. Holography maps a spin model with a finite dimensional Hilbert
space defined on a version of the stretched horizon into bulk gravitational
dynamics. The full Hamiltonian of the spin model contains a non-local
piece which generates chaotic dynamics, widely conjectured to be a
necessary part of quantum gravity, and a local piece which recovers
the perturbative spectrum in the bulk. 
\end{abstract}
\maketitle

\section{Introduction}

Previous work has argued for a unitary, holographic description of
black hole dynamics via certain spin models \citep{Lowe:2016mhi,Lowe:2017ehz}
defined on the stretched horizon \citep{thorne1986black} of the black
hole. These spin models have the common feature that non-local interactions
generate chaotic dynamics, widely conjectured to be an integral part
of a full quantum mechanical description of gravity \citep{Sekino:2008he}.
In this paper we argue that a similar approach works for the cosmological
horizon in de Sitter spacetime, given that the static patch metric
has a similar form to the metric in Schwarzschild coordinates. To
this end we will give an explicit prescription to map perturbative
bulk fields to a quantum mechanical operator defined in the holographic
spin model. This map will then allow us to construct a local Hamiltonian
that reproduces the classical energy of a perturbation around de Sitter
spacetime. We argue that this local Hamiltonian, together with the
non-local long-range interaction necessary to generate chaotic dynamics,
can potentially be a viable description of de Sitter quantum gravity.

Before we begin, we will review relevant facts of the de Sitter space-time
to establish our convention of notations. We mostly follow the conventions
in Ref.~\citep{Spradlin:2001pw}. Throughout the paper we will restrict
our discussion to $(1+3)$-dimensional space-time entirely, although
the methodology presented can in principle be applied to higher (or
lower) dimensions. Our metric signatures are always mostly positive,
i.e.~$(-++\cdots)$. 

\begin{figure}
\begin{tikzpicture}[scale=1]
    \draw (-2,2) -- (2,2);
    \draw (-2,2) -- (-2,-2);
    \draw (-2,-2) -- (2,-2);
    \draw (2,-2) -- (2,2);
    \draw[->,dashed] (-2,-2) -- (2,2);
    \draw[->,dashed] (2,-2) -- (-2,2);
    \shade (0,0) -- (2,2) -- (2,-2) -- (0,0);
    \pgfmathsetmacro{\e}{1.5}   % eccentricity
    \pgfmathsetmacro{\a}{0.7}
    \pgfmathsetmacro{\b}{(\a*sqrt((\e)^2-1)}
    \draw plot[domain=-1.66:1.66] ({\a*cosh(\x)+0.09},{\b*sinh(\x)});
  \end{tikzpicture} \caption{Penrose diagram for de Sitter spacetime, where shaded region is covered
by the static coordinates. The stretched horizon (solid curve inside
the static patch) is defined as a hypersurface at fixed $r$.}
\label{penrose}
\end{figure}
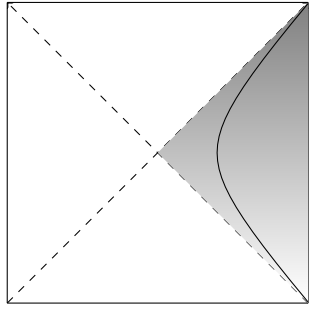

Static coordinates cover only one triangular region in the Penrose
diagram (see Fig.~\ref{penrose})
\begin{equation}
\dd s^{2}=-(1-\frac{r^{2}}{\ell^{2}})\,\dd t^{2}+\frac{\dd r^{2}}{1-\frac{r^{2}}{\ell^{2}}}+r^{2}\,\dd\Omega^{2}\label{eq:static}
\end{equation}
where $\dd\Omega^{2}=\dd\theta^{2}+\sin^{2}\theta\,\dd\varphi^{2}$
is the line element on the unit 2-sphere $S^{2}$ and $\ell$ is the
radius of curvature of the de Sitter spacetime.

\section{Mode Functions in De Sitter}

\subsection{Static patch}

Our goal is to construct a spin model which reproduces the bulk spectrum
in a de Sitter background. To proceed we first review some standard
results concerning mode functions in static de Sitter. For simplicity,
we will treat the massless minimally coupled scalar, and follow the
point of view of \citep{Danielsson:2002qh} that with a cutoff to
exclude the zero mode, the system can be quantized around the Bunch-Davies
(or Euclidean) vacuum state. Had we been interested in the system
including the zero-mode, this quantization preserving de Sitter isometries
would be inadmissible \citep{Allen:1987tz}. This quantization of
the massless minimally coupled scalar around the Bunch-Davies vacuum
is closely related to that of perturbative gravitons, as explained
in \citep{Bernar:2016zgq,Bernar:2018nww}. The results may then be
straightforwardly applied to other bulk modes once this case is understood.

We begin by considering modes in the static patch \eqref{eq:static}.
The equation of motion for the free scalar field $\Phi(t,r,\theta,\phi)$
is
\[
\frac{\partial_{t}^{2}\Phi}{1-r^{2}/\ell^{2}}-\frac{\partial_{r}[r^{2}(1-r^{2}/\ell^{2})\partial_{r}\Phi]}{r^{2}}-\frac{\partial_{\theta}(\sin\theta\partial_{\theta}\Phi)}{r^{2}\sin\theta}-\frac{\partial_{\varphi}^{2}\Phi}{r^{2}\sin^{2}\theta}=0\,.
\]
Separating variables, we have 
\[
\Phi_{\omega lm}(t,r,\theta,\varphi)=A_{\omega l}e^{-i\omega t}f_{\omega l}(r)Y_{lm}(\theta,\varphi)\,,
\]
where $f_{\omega l}(r)$ satisfies 
\[
(1-r^{2}/\ell^{2})f''_{\omega l}(r)+\frac{2(1-2r^{2}/\ell^{2})}{r}f'_{\omega l}(r)+\left(\frac{\omega^{2}}{1-r^{2}/\ell^{2}}-\frac{l(l+1)}{r^{2}}\right)f_{\omega l}(r)=0\,.
\]
We pick the set of solutions that are regular at $r=0$ and find 
\[
f_{\omega l}(r)=\frac{(r/\ell)^{l}}{\ell}(1-r^{2}/\ell^{2})^{i\omega\ell/2}{}_{2}{\rm F}_{1}\left(\frac{l+i\omega\ell}{2},\frac{l+i\omega\ell+3}{2};l+\frac{3}{2};\frac{r^{2}}{\ell^{2}}\right)\,.
\]
We fix the normalization constant $A_{\omega l}$ by computing the
Klein-Gordon norm. This is defined on a spacelike surface $\Sigma$
by
\begin{align*}
\left\langle f,g\right\rangle  & =-i\int_{\Sigma}\dd\Sigma\,n^{\lambda}\,(f\partial_{\lambda}g^{*}-g^{*}\partial_{\lambda}f)
\end{align*}
where $n^{\lambda}$ is a timelike unit vector normal to $\Sigma$.
Evaluating this on a constant $t$ slice gives
\begin{equation}
\left\langle f,g\right\rangle =-i\int(f\partial_{t}g^{\star}-g^{\star}\partial_{t}f)\frac{r^{2}\,\dd r\,\sin\theta\,\dd\theta\,\dd\varphi}{1-r^{2}/\ell^{2}}\,.\label{eq:innerprod}
\end{equation}
Computing the mode normalization then gives
\[
\left\langle \Phi_{\omega lm},\Phi_{\omega'l'm'}\right\rangle =A_{\omega l}A_{\omega'l}^{\star}(\omega+\omega')\delta_{ll'}\delta_{mm'}\int_{0}^{\ell}\frac{f_{\omega l}(r)f_{\omega'l}^{\star}(r)\,r^{2}\,\dd r}{1-r^{2}/\ell^{2}}\,.
\]
Using the equation of motion for $f_{\omega l}(r)$ we have 
\[
[r^{2}(1-r^{2}/\ell^{2})f'_{\omega l}(r)]'\,f_{\omega'l}^{\star}(r)+\left(\frac{\omega^{2}r^{2}}{1-r^{2}/\ell^{2}}-l(l+1)\right)f_{\omega l}(r)f_{\omega'l}^{\star}(r)=0
\]
and likewise 
\[
[r^{2}(1-r^{2}/\ell^{2})f_{\omega'l}^{\star\prime}(r)]'\,f_{\omega l}(r)+\left(\frac{\omega'^{2}r^{2}}{1-r^{2}/\ell^{2}}-l(l+1)\right)f_{\omega'l}^{\star}(r)f_{\omega l}(r)=0\,.
\]
Subtracting we have 
\[
\frac{(\omega^{2}-\omega'^{2})r^{2}}{1-r^{2}/\ell^{2}}f_{\omega l}(r)f_{\omega'l}^{\star}(r)=[r^{2}(1-r^{2}/\ell^{2})f_{\omega'l}^{\star\prime}(r)]'\,f_{\omega l}(r)-[r^{2}(1-r^{2}/\ell^{2})f'_{\omega l}(r)]'\,f_{\omega'l}^{\star}(r)\,,
\]
and integrating by parts, we have 
\[
\int_{0}^{\ell}\frac{(\omega^{2}-\omega'^{2})r^{2}}{1-r^{2}/\ell^{2}}f_{\omega l}(r)f_{\omega'l}^{\star}(r)=r^{2}(1-r^{2}/\ell^{2})f_{\omega'l}^{\star\prime}(r)\,f_{\omega l}(r)-r^{2}(1-r^{2}/\ell^{2})f'_{\omega l}(r)\,f_{\omega'l}^{\star}(r)\Big|_{0}^{\ell}\,.
\]
This gives 
\[
\int_{0}^{\ell}\frac{r^{2}\,\dd r}{1-r^{2}/\ell^{2}}f_{\omega l}(r)f_{\omega'l}^{\star}(r)=\frac{\ell^{2}}{\omega^{2}-\omega'^{2}}\lim_{r\to\ell}\,(1-r^{2}/\ell^{2})\left[f_{\omega'l}^{\star\prime}(r)\,f_{\omega l}(r)-f'_{\omega l}(r)\,f_{\omega'l}^{\star}(r)\right]\,.
\]
Using the hypergeometric identity near $z=1$ 
\[
_{2}{\rm F}_{1}(a,b,c,z)=\frac{\Gamma(c)\Gamma(c-a-b)}{\Gamma(c-a)\Gamma(c-b)}+\frac{\Gamma(c)\Gamma(a+b-c)}{\Gamma(a)\Gamma(b)}(1-z)^{c-a-b}
\]
we can expand $f_{\omega l}(r)$ near $r=\ell$ to give 
\[
\ell f_{\omega l}(r)\approx\frac{\Gamma(l+\frac{3}{2})\Gamma(-i\ell\omega)}{\Gamma\left(\frac{l-i\omega\ell}{2}\right)\Gamma\left(\frac{l-i\omega l+3}{2}\right)}(1-r^{2}/\ell^{2})^{\frac{i\ell\omega}{2}}+\frac{\Gamma(l+\frac{3}{2})\Gamma(i\ell\omega)}{\Gamma\left(\frac{l+i\omega\ell}{2}\right)\Gamma\left(\frac{l+i\omega l+3}{2}\right)}(1-r^{2}/\ell^{2})^{-\frac{i\ell\omega}{2}}\,.
\]
Letting 
\[
B_{\omega l}=\frac{\Gamma(l+\frac{3}{2})\Gamma(i\ell\omega)}{\Gamma\left(\frac{l+i\omega\ell}{2}\right)\Gamma\left(\frac{l+i\omega l+3}{2}\right)}
\]
we see that we have 
\[
\ell f_{\omega l}(r)\approx B_{\omega l}^{\star}(1-r^{2}/\ell^{2})^{\frac{i\ell\omega}{2}}+B_{\omega l}(1-r^{2}/\ell^{2})^{-\frac{i\ell\omega}{2}}
\]
and 
\[
f'_{\omega l}(r)\approx\frac{-i\omega}{1-r^{2}/\ell^{2}}\left[B_{\omega l}^{\star}(1-r^{2}/\ell^{2})^{\frac{i\ell\omega}{2}}-B_{\omega l}(1-r^{2}/\ell^{2})^{-\frac{i\ell\omega}{2}}\right]\,.
\]
Multiplying $f_{\omega l}(r)$ and $f'_{\omega l}(r)$ and dropping
terms that are rapidly oscillating as $|\omega-\omega'|>0$ and $r\to\ell$,
we find that 
\begin{equation}
\int_{0}^{\ell}\frac{r^{2}\,\dd r}{1-r^{2}/\ell^{2}}f_{\omega l}(r)f_{\omega'l}^{\star}(r)=\lim_{r\to\ell}\frac{2|B_{\omega l}|^{2}}{\omega-\omega'}\sin\left[\frac{\left(\omega-\omega'\right)\ell}{2}\log\left(\frac{1}{1-r^{2}/\ell^{2}}\right)\right]\label{eq:radialint}
\end{equation}
Using 
\[
\lim_{C\to\infty}\frac{\sin Cx}{x}=\pi\delta(x)
\]
we have 
\[
\int_{0}^{\ell}\frac{r^{2}\,\dd r}{1-r^{2}/\ell^{2}}f_{\omega l}(r)f_{\omega'l}^{\star}(r)=2\pi|B_{\omega l}|^{2}\delta(\omega-\omega')
\]
or 
\begin{equation}
\left\langle \Phi_{\omega lm},\Phi_{\omega'l'm'}\right\rangle =4\pi|A_{\omega l}|^{2}|B_{\omega l}|^{2}\delta_{ll'}\delta_{mm'}\omega\delta(\omega-\omega')\,.\label{eq:staticnorm}
\end{equation}
If we normalize according to
\[
\left\langle \Phi_{\omega lm},\Phi_{\omega'l'm'}\right\rangle =\delta_{ll'}\delta_{mm'}\omega\delta(\omega-\omega')\,,
\]
we need to pick $A_{\omega l}$ such that 
\[
|A_{\omega l}|^{2}=\frac{1}{4\pi|B_{\omega l}|^{2}}\,.
\]

\subsection{Flat slicing}

Now let us consider the analogous problem for modes in the flat slicing.
The metric takes the form
\[
\dd s^{2}=-\dd\tau^{2}+e^{2\tau/\ell}\dd\vec{x}^{2}=-\dd\tau^{2}+e^{2\tau/\ell}(\dd\rho^{2}+\rho^{2}\dd\theta^{2}+\rho^{2}\sin^{2}\theta\dd\varphi^{2})
\]
where $\tau\in(-\infty,+\infty)$ and $\rho\in(0,+\infty)$. The wave
equation for the massless minimally coupled scalar is given by 
\[
\partial_{\tau}^{2}\phi+\frac{3}{\ell}\partial_{\tau}\phi-e^{-2\tau/\ell}\Delta\phi=0
\]
where $\Delta\phi$ is the usual spatial Laplacian operator 
\[
\Delta\phi=\partial_{\rho}^{2}\phi+\frac{2}{\rho}\partial_{\rho}\phi+\frac{\partial_{\theta}^{2}\phi}{\rho^{2}}+\frac{\partial_{\theta}\phi}{\rho^{2}\tan\theta}+\frac{\partial_{\varphi}^{2}\phi}{\rho^{2}\sin^{2}\theta}\,.
\]
Separating variables, we use the ansatz 
\[
\phi_{klm}(\tau,\rho,\theta,\varphi)=T_{k}(\tau)R_{kl}(\rho)Y_{lm}(\theta,\varphi)
\]
where $T_{k}(\tau)$ satisfies 
\[
T''(\tau)+\frac{3}{\ell}T'(\tau)+k^{2}e^{-2\tau/\ell}T(\tau)=0
\]
and $R_{kl}(\rho)$ satisfies 
\[
R''(\rho)+\frac{2}{\rho}R'(\rho)+\left(k^{2}-\frac{l(l+1)}{\rho^{2}}\right)R(\rho)=0\,.
\]
Assuming regularity at $\rho=0$ we can solve for $R(\rho)$
\[
R_{kl}(\rho)=C_{kl}\,j_{l}(k\rho)
\]
where $j_{l}(x)$ is the spherical Bessel function of first kind $j_{l}(x)=\sqrt{\frac{\pi}{2x}}J_{l+\frac{1}{2}}(x)$.
We will determine the normalization constant $C_{kl}$ later.

The equation for $T(\tau)$ can be solved to give 
\[
T_{k}(\tau)=c_{1}e^{ik\ell e^{-\tau/\ell}}\left(1-ik\ell e^{-\tau/\ell}\right)+c_{2}e^{-ik\ell e^{-\tau/\ell}}\left(1+ik\ell e^{-\tau/\ell}\right)
\]
or, in terms of the conformal time $\eta=-\ell e^{-\tau/\ell}$ 
\[
T_{k}(\tau)=c_{1}e^{-ik\eta}\left(1+ik\eta\right)+c_{2}e^{ik\eta}\left(1-ik\eta\right)\,.
\]
We assume Bunch-Davies vacuum and therefore pick the special solution
\[
T_{k}(\eta)=e^{-ik\eta}\left(1+ik\eta\right)
\]
and absorb the normalization constant into $C_{kl}$, which we will
fix now. The mode functions are normalized according to the Klein-Gordon
norm 
\[
\left\langle f,g\right\rangle =-i\int_{\Sigma}(f\partial_{\mu}g^{\star}-g^{\star}\partial_{\mu}f)\,n^{\mu}\sqrt{\gamma}\,\dd^{3}x\,.
\]
Here $\Sigma$ is a spacelike hypersurface with unit norm $n^{\mu}$
and $\sqrt{\gamma}$ is the spatial volume element. We pick the $\tau=0$
timeslice in the flat slicing since the metric at $\tau=0$ is conveniently
the Minkowski metric. In addition, we have $\partial_{\tau}=\partial_{\eta}$
on the $\tau=0$ timeslice. We therefore have 
\[
\left\langle f,g\right\rangle =-i\int(f\partial_{\tau}g^{\star}-g^{\star}\partial_{\tau}f)\,\rho^{2}\,\dd\rho\,\sin\theta\,\dd\theta\,\dd\varphi\,.
\]
We use this to fix the normalization factor $C_{kl}$. We have, for
the $\phi_{klm}$ modes on $\tau=0$ 
\[
\partial_{\tau}\phi_{klm}=-C_{kl}\,k^{2}\ell\,e^{ik\ell}j_{l}(k\rho)\,Y_{lm}(\theta,\varphi)\,.
\]
Therefore 
\[
\left\langle \phi_{klm},\phi_{k'l'm'}\right\rangle =i\ell C_{kl}C_{k'l'}^{\star}\,\delta_{ll'}\delta_{mm'}e^{i(k-k')\ell}\int[(1-ik\ell)k'^{2}-(1+ik'\ell)k^{2}]j_{l}(k\rho)j_{l}(k'\rho)\rho^{2}\,\dd\rho\,.
\]
Using the orthogonality of spherical Bessel functions 
\[
\int_{0}^{\infty}\rho^{2}j_{l}(u\rho)j_{l}(v\rho)\,\dd\rho=\frac{\pi}{2u^{2}}\delta(u-v)
\]
we have 
\[
\left\langle \phi_{klm},\phi_{k'l'm'}\right\rangle =\ell^{3}\,\pi k\,|C_{kl}|^{2}\,\delta_{ll'}\delta_{mm'}\delta(k\ell-k'\ell)\,.
\]
We therefore find 
\[
C_{kl}=\frac{1}{\sqrt{\pi k\ell}}\frac{1}{\ell}\,,
\]
and the full solution is therefore 
\[
\phi_{klm}(\tau,\rho,\theta,\varphi)=\frac{1}{\ell\sqrt{\pi k\ell}}\,e^{-ik\eta}(1+ik\eta)\,j_{l}(k\rho)\,Y_{lm}(\theta,\varphi)\,.
\]

\subsection{Matching modes across the cosmological horizon}

\begin{figure}

\includegraphics{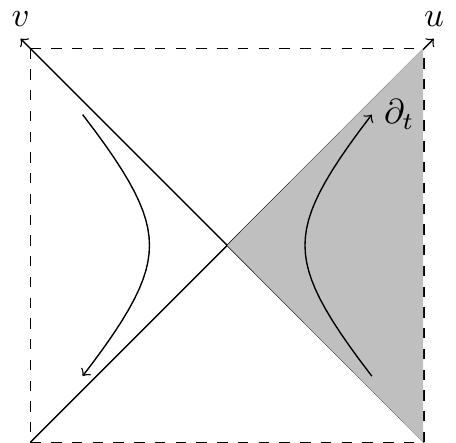}\caption{\label{fig:Penrose-diagram-for}Penrose diagram for de Sitter. Flat
slicing modes cover the right upper half of the diagram and are matched
to static patch modes on the line $u=0$. Positive frequency modes
in the Bunch-Davies vacuum are then analytic in the lower-half-complex
$v-$plane.}

\end{figure}

Following \citep{Unruh}, modes in the flat slicing may be viewed
as modes entangled across the left and right static patches. In Kruskal
coordinates in the right static patch $u=e^{x^{-}/\ell}$ and $v=-e^{-x^{+}/\ell}$
(other choices of sign generate the other patches in static coordinates)
where $x^{\pm}=t\pm r^{\star}$ and 
\[
r^{\star}=\frac{\ell}{2}\log\frac{1+r/\ell}{1-r/\ell}\approx\frac{\ell}{2}\log\frac{2}{1-r/\ell}
\]
we have 
\[
\Phi_{\omega lm}\approx A_{\omega l}(B_{\omega l}^{\star}2^{i\omega\ell}|v|^{i\omega\ell}+B_{\omega l}2^{-i\omega\ell}|u|^{-i\omega\ell})Y_{lm}(\theta,\varphi)
\]
 near the cosmological horizon.

We now define a mode function $\Phi_{\omega lm}^{+}$ which is non-zero
on the right quadrant, and a $\Phi_{\omega lm}^{-}$ which is non-zero
on the left quadrant. We require that the linear combination 
\[
\bar{\Phi}_{\omega lm}=c\,\Phi_{\omega lm}^{+}+d\,\Phi_{\omega lm}^{-}
\]
be analytic in the lower half $v-$plane on the past horizon on the
right (and the past horizon on the left), i.e. the surface $u=0$.
On the right quadrant we can rewrite the function $(-v)^{i\omega\ell}=(e^{i\pi}v)^{i\omega\ell}=e^{-\pi\omega\ell}v^{i\omega\ell}$.
Therefore, the linear combination $\Phi_{\omega lm}^{+}+e^{-\pi\omega\ell}\Phi_{\omega lm}^{-}$
is analytic in the lower-half-complex $v-$plane at $u=0$ (see fig.
\ref{fig:Penrose-diagram-for}) corresponding to a combination of
positive frequency flat-slicing modes. The properly normalized mode
is 
\begin{equation}
\bar{\Phi}_{\omega lm}=\frac{1}{\sqrt{2\sinh(\pi\omega\ell)}}\left(e^{\pi\omega\ell/2}\Phi_{\omega lm}^{+}+e^{-\pi\omega\ell/2}\Phi_{\omega lm}^{-}\right)\,.\label{eq:entmode}
\end{equation}
This mode is analytic in the lower-half $v$-plane for either choice
of sign for $\omega$, so should be identified with a linear combination
of positive $k$ flat-slicing modes.

We can now expand a quantum field operator $\hat{\Phi}$ in terms
of normal modes 
\begin{align*}
\hat{\Phi} & =\int_{0}^{\infty}dk\sum_{lm}\phi_{klm}a_{klm}+{\rm h.c.\,(flat\,slicing)}\\
 & =\int_{0}^{\infty}d\omega\sum_{lm}\Phi_{\omega lm}^{+}b_{\omega lm}^{+}+\Phi_{\omega lm}^{{\rm -}}b_{\omega lm}^{{\rm -}\dagger}+{\rm h.c.\,(static\,patches)}\\
 & =\int_{-\infty}^{\infty}d\omega\sum_{lm}\bar{\Phi}_{\omega lm}c_{\omega lm}+{\rm h.c.\,(entangled\,static\,patches)}\,.
\end{align*}
We define the Bunch-Davies vacuum $|0\rangle$ to be annihilated by
all $a_{klm}$ with $k>0$
\[
a_{klm}|0\rangle=0
\]
which coincides with
\[
c_{\omega lm}\left|0\right\rangle =0
\]
for $-\infty<\omega<\infty$. We similarly define the static patch
vacuum $|\Omega\rangle$ to be annihilated by all $b_{\omega lm}^{\pm}$
with $\omega>0$
\[
b_{\omega lm}^{\pm}|\Omega\rangle=0\,.
\]
From the relation we obtained between $\Phi^{\pm}$ and $\bar{\Phi}$
we obtain the relations between $b_{\omega lm}^{\pm}$ and $c_{\omega lm}$
\begin{align*}
c_{\omega lm} & =\begin{cases}
\frac{1}{\sqrt{2\sinh{\pi\omega\ell}}}\left(e^{\pi\omega\ell/2}b_{\omega lm}^{+}+e^{-\pi\omega\ell/2}b_{\omega lm}^{-\dagger}\right) & \omega>0\\
\frac{1}{\sqrt{2\sinh{\pi\omega\ell}}}\left(e^{\pi\omega\ell/2}b_{\omega lm}^{+\dagger}+e^{-\pi\omega\ell/2}b_{\omega lm}^{-}\right) & \omega<0\,.
\end{cases}
\end{align*}
As usual, the vacuum state $\left|0\right\rangle $ becomes a thermal
density matrix in the right static patch when the modes $b_{\omega lm}^{-}$
are traced over.

We now need to compute the overlap of modes $\Phi_{\omega lm}$ with
$\phi_{\omega lm}$ to construct the Bogoliubov transformation. In
order to perform the integrals, we need the coordinate transformation
between $(t,r)$ and $(\eta,\rho)$. We define null coordinates in
the flat slicing
\[
U=\frac{\eta-\rho}{2}\qquad V=\frac{\eta+\rho}{2}
\]
leading to the metric. In the flat slicing using the $(\eta,\rho)$
coordinates the metric is 
\[
\dd s^{2}=\frac{\ell^{2}}{\eta^{2}}(-\dd\eta^{2}+\dd\rho^{2}+\rho^{2}\dd\Omega^{2})
\]
which in terms of $(U,V)$ coordinates becomes 
\[
\dd s^{2}=\frac{\ell^{2}}{(U+V)^{2}}(-4\dd U\dd V+(V-U)^{2}\dd\Omega^{2})\,.
\]
In the static patch, we define null coordinates $(u,v)$ 
\[
u=e^{x^{-}/\ell}\,,v=-e^{-x^{+}/\ell}
\]
and one verifies that the metric is 
\[
\dd s^{2}=\frac{\ell^{2}}{(1-uv)^{2}}(-4\dd u\dd v+(1+uv)^{2}\dd\Omega^{2})\,.
\]
We see that the relation between $(u,v)$ and $(U,V)$ is simply 
\[
U=-\frac{\ell}{u}\qquad V=\ell v
\]
which gives 
\[
\eta=-\frac{\ell}{u}+\ell v\qquad\rho=\frac{\ell}{u}+\ell v\,.
\]
The flat slicing mode function therefore becomes 
\[
\phi_{klm}=\frac{1}{\ell}\frac{1}{\sqrt{\pi k\ell}}\,e^{ik\ell(\frac{1}{u}-v)}\left[1-ik\ell\left(\frac{1}{u}-v\right)\right]\,\frac{\sin(k\ell(v+1/u)-l\pi/2)}{k\ell(v+1/u)}\,Y_{lm}(\theta,\varphi)
\]
where we have used the behavior of spherical Bessel function at infinity
\[
\lim_{\rho\to\infty}j_{l}(k\rho)=\frac{\sin(k\rho-l\pi/2)}{k\rho}\,.
\]
Near the past horizon on the left patch, $u\to0$ with $v$ fixed,
the flat slice mode becomes 
\[
\phi_{klm}=\frac{1}{\ell}\frac{-i}{\sqrt{\pi k\ell}}\,e^{ik\ell(\frac{1}{u}-v)}\sin(k\ell(v+1/u)-l\pi/2)\,Y_{lm}(\theta,\varphi)\,.
\]
Using the identity $\sin z=(e^{iz}-e^{-iz})/2i$ we can rewrite the
flat slice mode function as 
\begin{align*}
\phi_{klm} & =\frac{1}{\ell}\frac{-1}{2\sqrt{\pi k\ell}}\,e^{ik\ell(\frac{1}{u}-v)}(i^{-l}e^{ik\ell(v+1/u)}-i^{l}e^{-ik\ell(v+1/u)})Y_{lm}(\theta,\varphi)\\
 & =\frac{1}{\ell}\frac{1}{2\sqrt{\pi k\ell}}(i^{l}e^{-2ik\ell v}-i^{-l}e^{2ik\ell/u})Y_{lm}(\theta,\varphi)\sim\frac{1}{\ell}\frac{i^{l}}{2\sqrt{\pi k\ell}}e^{-2ik\ell v}Y_{lm}(\theta,\varphi)\,.
\end{align*}
We have shown above that near the past horizon the static patch mode
is 
\[
\Phi_{\omega lm}=A_{\omega l}B_{\omega l}^{\star}(-v)^{i\omega\ell}2^{i\omega\ell}Y_{lm}(\theta,\varphi)\,.
\]
On the past horizon the Klein-Gordon norm is 
\[
\left\langle f,g\right\rangle =-i\ell^{2}\int\dd\Omega\dd v\,(f\partial_{v}g^{\star}-g^{\star}\partial_{v}f)=i\ell^{2}\int_{-\infty}^{0}\dd\Omega\left(2\int g^{\star}\partial_{v}f\,\dd v-g^{\star}f\Big|_{0}^{-\infty}\right)\,.
\]
Since we have 
\[
\partial_{v}\Phi_{\omega lm}=i\omega\ell v^{-1}\Phi_{\omega lm}
\]
adding a small imaginary part to $\omega$ and $k$ to dampen the
oscillation of $(-v)^{i\omega\ell}$ and $e^{-ikv}$ we have, since
the boundary term vanishes, 
\[
\left\langle \Phi_{\omega lm},\phi_{klm}\right\rangle =\frac{(-i)^{l}}{\sqrt{\pi k\ell}}A_{\omega l}B_{\omega l}^{\star}2^{i\omega\ell}\int_{0}^{+\infty}e^{-2ik\ell v}v^{i\omega\ell-1}\,\dd v
\]
where we replaced $v\to-v$. This can be evaluated to give 
\[
\left\langle \Phi_{\omega lm},\phi_{klm}\right\rangle =\frac{(-i)^{l}}{\sqrt{\pi k\ell}}A_{\omega l}B_{\omega l}^{\star}(ik\ell)^{-i\omega\ell}\Gamma(i\omega\ell)\,.
\]
By choosing the phase of $A_{\omega l}$ appropriately such that 
\[
A_{\omega l}B_{\omega l}^{\star}=\frac{1}{\sqrt{4\pi}}
\]
we finally have 
\[
\left\langle \Phi_{\omega lm},\phi_{klm}\right\rangle =\frac{(-i)^{l}}{\sqrt{k\ell}}\frac{(ik\ell)^{-i\omega\ell}\Gamma(i\omega\ell)}{2\pi}\,.
\]
With the Bogoliubov transformation at hand, we can now map modes in
the flat slicing to entangled modes in the left and right static patch.
These in turn define modes in the upper quadrant of the static slicing
by continuation.

\section{Holographic Map}

Our goal is to build a holographic version of the bulk theory that
might be viewed as living on the so-called ``stretched horizon''
\citep{thorne1986black}. The essence of the black hole membrane paradigm
is that to an external observer outside the horizon, the black hole
horizon behaves more or less like a hydrodynamic membrane with properties
such as resistance and viscosity. Quantum mechanically, the stretched
horizon acts as a mirror \citep{Hayden:2007cs} which scrambles and
reflects information sent into it. Our viewpoint in this paper is
that since the static patch of de-Sitter has essentially the same
mathematical form as the Schwarzschild metric, one ought to be able
to construct a similar stretched horizon theory for static de-Sitter.
We further assume that the horizon entropy of de Sitter is to be matched
with the logarithm of the Hilbert space dimension. Thus, the stretched
horizon theory will be a finite dimensional quantum mechanical system.

Usually the stretched horizon is defined as the constant $r$ surface
such that the local temperature measured by a fiducial observer (constant
$r$) is the Planck temperature. When redshifted down to $r=0$, this
will match the Hawking temperature. In other words, one usually defines
$r_{\star}$ such that 
\begin{equation}
\frac{1}{2\pi\ell_{p}}\sqrt{1-r_{\star}^{2}/\ell^{2}}=T_{{\rm H}}=\frac{1}{2\pi\ell}\,,\label{eq:usualrs}
\end{equation}
where $\ell_{p}$ is the Planck length.

As it stands, the bulk Hilbert space is infinite dimensional, labelled
by the oscillators $c_{\omega lm}$ where $\omega\in\mathbb{R}$,
and the angular momentum ranges up to infinity. Each oscillator mode
creates a mode entangled across both static patches, with a stress-energy
tensor non-singular on the future cosmological horizon on the right
patch.

As a first step, we can discretize the sphere in coordinate space.
There are many possibilities for such a discretization, and the details
will not be too important for us, except to note that such a discretization
will produce an effective cutoff $l_{max}$ on the angular momentum.

Likewise, it is necessary to discretize the frequency $\omega$ which
can in turn be viewed as a radial quantum number. This discretization
may then be viewed as a kind of regulator for the radial coordinate.
In order to produce a useful effective field theory with such a cutoff
we choose a finite set of frequencies in the range
\begin{equation}
\frac{1}{\ell_{p}n_{UV}}>|\omega|\geq\frac{\pi}{\ell\log\left(\ell/\ell_{p}\right)}\,.\label{eq:iruv}
\end{equation}
For simplicity we can take the $\omega's$ to be evenly spaced in
this range, with spacing $\pi/\ell\log\left(\ell/\ell_{p}\right)$.
This corresponds to $n_{rad}=2\ell\log(\ell/\ell_{p})/\pi\ell_{p}n_{UV}$
radial points in the static patch. In particular, this number is conserved
with time. Here $n_{UV}>1$ is a factor introduced to parameterize
the ultraviolet cutoff. We will see momentarily why the $\log$ factor
appears.

An issue we immediately face is regulating the continuum mode normalization
\eqref{eq:staticnorm} to the discrete case. To do this we replace
the upper limit on the radial integral in \eqref{eq:innerprod} by
$\ell\to\ell-\epsilon.$ Then the result of the integral \eqref{eq:radialint}
may be replaced by
\begin{align*}
\int_{0}^{\ell-\epsilon}\frac{r^{2}\,\dd r}{1-r^{2}/\ell^{2}}f_{\omega l}(r)f_{\omega'l}^{\star}(r) & =\frac{2|B_{\omega l}|^{2}}{\omega-\omega'}\left.\sin\left[\frac{\left(\omega-\omega'\right)\ell}{2}\log\left(\frac{1}{1-r^{2}/\ell^{2}}\right)\right]\right|_{r=\ell-\epsilon}\,.
\end{align*}
Keeping in mind $\omega-\omega'=\pi n/\ell\log\left(\ell/\ell_{p}\right)$
for some integer $n$ we choose
\begin{equation}
\left.\log\left(\frac{1}{1-r^{2}/\ell^{2}}\right)\right|_{r=\ell-\epsilon}=2\log\left(\ell/\ell_{p}\right)\label{eq:horpos}
\end{equation}
which fixes $r_{*}$ according to \eqref{eq:usualrs},
\[
\int_{0}^{\ell-\epsilon}\frac{r^{2}\,\dd r}{1-r^{2}/\ell^{2}}f_{\omega l}(r)f_{\omega'l}^{\star}(r)=2\ell\log\left(\frac{\ell}{\ell_{p}}\right)|B_{\omega l}|^{2}\delta_{\omega,\omega'}
\]
up to rapidly oscillating terms. This unusual relation between a short
distance cutoff and an infrared cutoff is typical in holographic models.

Finally, each harmonic oscillator mode $c_{\omega lm}$ produces an
infinite dimensional Hilbert space. To regulate these Hilbert subspaces,
we use the Holstein-Primakoff map \citep{holstein} to replace $c_{\omega lm}$
by spin operators, introducing the parameter $s_{max}\gg1$
\[
s_{+}^{\omega lm}=\sqrt{2s_{max}}\sqrt{1-\frac{c_{\omega lm}^{\dagger}c_{\omega lm}}{2s_{max}}}c_{\omega lm},\,s_{-}^{\omega lm}=\sqrt{2s_{max}}c_{\omega lm}^{\dagger}\sqrt{1-\frac{c_{\omega lm}^{\dagger}c_{\omega lm}}{2s_{max}}},\,s_{z}^{\omega lm}=s-c_{\omega lm}^{\dagger}c_{\omega lm}\,.
\]
 For states near the ground state, we can approximate $\sqrt{1-\frac{c_{\omega lm}^{\dagger}c_{\omega lm}}{2s_{max}}}$
by $1$.

This regularization of the Hilbert space then allows us to write the
energy in the scalar field at quadratic order as a spin model
\[
H_{0}=\sum_{\left\{ \omega\right\} }\sum_{l=0}^{l_{max}}\sum_{m=-l}^{l}\omega\left(c_{\omega lm}^{\dagger}c_{\omega lm}+c_{\omega lm}c_{\omega lm}^{\dagger}\right)\,.
\]
The dimension of the Hilbert space, for large $l_{max}$ is $(2s_{max}+1)^{l_{max}^{2}n_{rad}}=e^{S_{BH}}$,
identified with the Bekenstein-Hawking entropy of the cosmological
horizon $S_{BH}=\pi\ell^{2}/\ell_{p}^{2}\equiv N$. If we follow the
arguments of \citep{cohen}, we identify
\[
l_{max}^{2}n_{rad}\log s_{max}\sim N
\]
and a natural choice would be to scale $n_{rad}\sim l_{max}\sim N^{1/3}$,
dropping subleading $\log$ factors for simplicity in a large $N$
limit. This leads to a short distance cutoff length of order $\ell_{p}N^{1/6}$
in all directions (and a choice $n_{UV}\sim N^{1/6}$). We note if
our present universe was replaced by a pure de Sitter region with
the same Hubble parameter, we would find $N\approx10^{120}$ and $\ell_{p}N^{1/6}$
would correspond to a $GeV$ UV cutoff.

\begin{comment}
If we wish, we can pass from angular momentum modes to modes local
on the discretized sphere by doing $c_{\omega lm}=\int d\Omega Y_{lm}^{*}c_{\omega}(\theta,\phi)$
and using $\sum_{l,m}Y_{lm}(\theta,\phi)Y_{lm}^{*}(\theta',\phi')=\frac{1}{\sin\theta}\delta(\theta-\theta')\delta(\phi-\phi')$
but will generate some nontrivial kernel.
\end{comment}

So far, we have simply regulated the scalar field theory at the level
of free field theory and found a holographic dual that reproduces
that. The holographic dual can be viewed as living on an $S^{2}$
with the discrete parameter $\omega$ labelling different variables
at each point on the sphere. This construction is guaranteed to reproduce
the bulk correlators of free scalar field theory with this particular
regulator.

The ground state of the Hamiltonian corresponds to the Bunch-Davies
vacuum state, and the Hamiltonian is diagonal in modes that are entangled
between the left and right ``patches''. Tracing over one set leads
to an approximately thermal density matrix in the other, subject to
the regulator on mode number imposed by finite $s_{max}$. The excitations
of this model will lead to stress-energy tensors regular on the cosmological
horizon, avoiding the firewall conundrum.

In general, we also expect to have to add perturbative interactions
to this model, which will typically be suppressed by powers of $N$
relative to the quadratic term. One might hope to follow a construction
paralleling HKLL \citep{Hamilton:2005ju,Hamilton:2006az} to reproduce
perturbative field theory in the bulk.

Such a theory might be satisfactory for de Sitter spacetime. Once
the initial state corresponding to Bunch-Davies is specified on the
past horizon of the right static patch (and its continuation onto
the left static patch) it evolves according to the standard rules
of quantum mechanics. The future cosmological horizon would essentially
behave like a remnant, becoming entangled with the degrees of freedom
in the left patch. A priori this poses no issues for the information
problem, because the cosmological horizon in de Sitter is eternal.

Motivated by the physics of black hole horizons, it is interesting
to explore what happens when this model is supplemented by an additional
nonlocal term as studied in \citep{Lowe:2016mhi,Lowe:2017ehz,Lowe:2019scv}
which is thought to generate chaotic dynamics over sufficiently long
timescales. In the black hole case, the timescale associated with
quantum scrambling is linked to the timescale the horizon can retain
quantum information, before emitting it to the region outside the
black hole. In the de Sitter case, we view the static patch as analogous
to the black hole interior and are mostly interested in developing
the holographic map on timescales shorter than this scrambling time.
We may then study the decoherence of local observables built using
the holographic map described above, when supplemented by chaotic
interactions.

The full Hamiltonian includes a non-local piece and a local piece,
where the non-local piece is given by 
\begin{equation}
H_{\textrm{nl}}=\sum_{ijkl}J_{ijkl}s_{i}s_{j}s_{k}s_{l}\label{eq:nonlocalham}
\end{equation}
Here the coupling $J_{ijkl}$ is drawn randomly from a Gaussian distribution
with zero mean (tensor indices are suppressed). We do not have in
mind averaging over this coupling, but rather work with a fixed set
of $J_{ijkl}$ as needed to generate chaotic dynamics. We impose the
condition that the variance of the non-local Hamiltonian $\var(H_{{\rm nl}})=1$.
This forces the width of the Gaussians to scale like $1/N^{2}$, due
to the following analysis 
\begin{equation}
1=\langle H_{nl}^{2}\rangle\sim J^{2}\left\langle \sum_{i_{1}\cdots i_{8}}s_{i_{1}}\cdots s_{i_{8}}\right\rangle \sim J^{2}N^{4}\label{eq:scaling}
\end{equation}
where in the last step we have used the fact that on average $\langle s_{i}s_{j}\rangle=\delta_{ij}$.
We note this unusual scaling with $N$ is designed to reproduce the
Bekenstein-Hawking entropy via microstate counting for fixed $N$
as opposed to the more conventional large $N$ limit where $\left\langle H_{nl}^{2}\right\rangle \sim N$,
which would widen the spectrum to much larger energies.

Our proposal for the full Hamiltonian is then
\begin{equation}
H=H_{0}+T_{H}H_{nl}\label{eq:hamiltonian}
\end{equation}
and the chaotic term may then be treated as a small perturbation for
short enough time intervals, where it will shift energies at leading
order by terms of order $T_{H}$.

One may then study how local perturbations of the thermal state decohere
when this term is included. Following the analysis of \citep{Lowe:2019scv}
we expect the timescale of such decoherence to be
\begin{equation}
t_{dec}=\beta\log N\label{eq:deco}
\end{equation}
This resembles the scrambling time, however the interpretation here
is somewhat different. With the scaling \eqref{eq:scaling} the global
scrambling time is expected to be
\[
t_{scr}=\beta N^{1/2}\log N\gg t_{dec}
\]
if the bounds derived in \citep{Bentsen:2018uph} happen to be saturated.
However, the local decoherence time is the quantity of most relevance
in deciding when the holographic map derived above breaks down. A
similar breakdown of the bulk description via effective field theory
in a black hole interior was noted in \citep{Lowe:2015eba,Lowe:2016mhi,Lowe:2017ehz}.

Given that a local operator will evolve to a highly non-local operator
in the time \eqref{eq:deco}, rather than simply undergoing the free
propagation governed by the term $H_{0}$, our holographic map based
on the mode functions \eqref{eq:entmode} will break down after this
timescale. In the case of applying this to a pure de Sitter region
with $\ell$ matched to our present cosmological horizon, this would
imply a breakdown in the local laws of physics after a timescale of
order 4000 billion years due to quantum gravity effects. It would
be very interesting to devise experiments sensitive to this local
decoherence. While the shifts in energy levels are tiny, of order
$10^{-33}eV$ the nonlocal character of the decoherence opens the
door to more sensitive experiments.

One might wonder whether such a holographic description is ruled out
for primordial inflation. In that case, one can try to embed ``small''
de Sitter models into a much larger Hilbert space, which is needed
to describe the late-time phase of cosmology. Holographic bounds with
these considerations in mind were considered in \citep{Banks:2003pt,Lowe:2004zs}.
The decoherence times in this case can be made much longer than the
timescale associated with primordial inflation.

It should also be noted that once a local basis of operators has decohered,
for example in Heisenberg picture
\[
c_{\omega lm}(t)=e^{iHt}c_{\omega lm}(0)e^{-iHt}
\]
with $t>t_{dec}$ one may simply do a change of basis by the unitary
transformation $e^{iHt_{dec}}$ to return to another local basis
\[
\tilde{c}_{\omega lm}(t)=e^{-iHt_{dec}}c_{\omega lm}(t)e^{iHt_{dec}}
\]
therefore, in some basis, one always retains an approximately local
description of spacetime physics. This realizes the proposal of \citep{Lowe:2015eba}
in a concrete model, when adapted to de Sitter spacetime.

\section{Conclusions}

Now that we have a detailed proposal for the stretched horizon theory
of the de Sitter cosmological horizon, we can try to adapt the method
to black holes. A key step in the development of the holographic map
was the assumption of regularity of the modes on the pole of the static
patch. This eliminated the non-normalizable modes and allowed us to
make a one-one map from frequency/radial quantum number space to mode
functions \eqref{eq:entmode}. For black holes in asymptotically flat
space, one would need to perform a similar restriction, which might
be accomplished by placing a mirror around the black hole to prevent
evaporation. In practice, as we have learned over the years, the best
substitute for this procedure is simply to introduce a negative cosmological
constant which has the same effect and can be handled much more precisely.
Thus, we expect the present considerations will apply largely unchanged
to a large black hole in anti-de Sitter spacetime which does not evaporate.
In this way, we can use the present construction to derive a holographic
map for the interior of such a black hole. One might then hope to
derive the spin model directly from the conformal field theory description
available in that case. Note here we have in mind realizing the black
hole in a single conformal field theory representing, perhaps, a large
black hole formed by collapse, rather than the tensor product conformal
field theories describing wormholes.

Turning this argument around, we then expect the much more interesting
case of the evaporating black hole in asymptotically flat space, or
a small black hole in asymptotically anti de Sitter space will involve
important extra ingredients. The coupling between this stretched horizon
theory and some larger holographic theory describing the asymptotic
region will need to be specified. Nevertheless, for timescales shorter
than $t_{dec}$ we expect to be able to apply the considerations of
the present paper, which is sufficient to extend the holographic map
to black hole interiors.

In the case of anti-de Sitter/conformal field theory duality, it is
often suggested one has control of the holographic map all the way
to the stretched horizon. In that case one has a fixed local basis
extending from asymptotic infinity down to the stretched horizon.
The present picture implies the coupling between the exterior and
the stretched horizon eventually become highly nonlocal, contaminating
the exterior physics with non-local effects. Indeed, nonlocal interactions
akin to \eqref{eq:nonlocalham} must emerge from the correspondence
in a smooth way as one approaches the stretched horizon. This has
the profound consequence that nonlocal scrambling effects might be
detected outside large black holes, if sufficiently long timescales
can be probed to overcome the $T_{H}$ suppression factor in \eqref{eq:hamiltonian}.
Indeed, such effects are probed in current gravitational wave experiments
\citep{LIGOScientific:2018mvr}. For example, for black hole mergers
with masses of order a solar mass, one must probe around $100$ light
crossing times to access the timescale \eqref{eq:deco}. As these
experiments become more precise it will be very interesting to look
for signs of violations of the equivalence principle. For example,
one might look for anomalies in the late time ringing profile following
black hole merger.

Finally, we end with a comment on an interesting numerological coincidence
of this holographic model. We noted above, that if we replace our
present cosmology with a de Sitter horizon with size around $14$
billion light years, an unacceptably small ultraviolet cutoff emerges
on bulk effectively field theory of about $1$ GeV. This may simply
be a signal that a more precise holographic model would produce a
bulk cutoff in a much more subtle way. However, for now, let us instead
explore the possibility that the current observable entropy $S\approx10^{88}$
which arises largely from cosmic microwave background photons, might
be equated with a late-time de Sitter entropy. Interestingly, this
predicts the cosmic acceleration must increase versus the previous
possibility, a feature also noted in the Hubble tension experiments,
and the ultraviolet cutoff that emerges is the more experimentally
interesting value of $100$ TeV. This raises the possibility that
holographic physics might appear in collider experiments at experimentally
accessible scales. Unfortunately, the model also predicts the horizon
size must shrink to of order $10^{4}$ m to reach the late time de
Sitter phase, so we are presently far off from the phase, and it is
not clear how much to trust the ultraviolet cutoff result. Nevertheless,
the model was designed so a freely falling observer will use a cutoff
with fixed proper spatial resolution, so there is reason to be optimistic.
\begin{acknowledgments}
D.L. is supported in part by DOE grant de-sc0010010. 
\end{acknowledgments}

\bibliographystyle{utphys}
\bibliography{ref}

\end{document}